\documentclass[10pt,reqno]{amsart}
\usepackage{graphicx}
\usepackage{indentfirst,csquotes}
\usepackage{xcolor,paralist,hyperref,titlesec,fancyhdr,etoolbox}
\usepackage{secdot}
\usepackage{amssymb,amsthm,amsmath}
\usepackage{xcolor,paralist,hyperref,titlesec,fancyhdr,etoolbox}

\titleformat{\section}[display]{\normalfont\huge\bfseries\centering}{\centering}{10pt}{\Large}

\hypersetup{ colorlinks=true, linkcolor=black, filecolor=black, urlcolor=black }

\usepackage{lipsum}

\begin{document}
\title{Leggett-Garg test inequality with spin and flavour neutrino oscillations in a constant magnetic field} 
\author{I. A. Monroy, S. D. Madrid\\Universidad Distrital Francisco Jos\'e de Caldas}
\date{}


\begin{abstract}
The Leggett-Garg inequality (LGI), an analogue of Bell’s inequality involving correlations of measurements of one observable on a system at different times, stands as one of the hallmark tests of quantum mechanics against classical predictions. In this work, we investigate its implications in the context of neutrino flavour ($\nu_{e}^{L}\leftrightarrow\nu_{\mu}^{L}$) and spin ($\nu_{e}^{L}\leftrightarrow\nu_{e}^{R}$) oscillations in the presence of a constant transverse magnetic field. For systems with strong magnetic fields, we show that, for both cases, there are length regions $\Delta L$ where the LGI is violated, as quantified by the correlator functions $K_3$ and $K_4$. 
\end{abstract} 
\maketitle
\section{Introduction}
\label{intro}
Quantum mechanics describes microscopic systems in terms of the superposition of distinct states associated with the spectra of physical observables. When one measurement of these physical observables is performed on a quantum system, it yields one single outcome, with some probability given by quantum mechanics. This idea highlights the distinction between classical systems and the quantum behavior of microscopic systems, as tested through Bell’s and Leggett–Garg inequalities. Bell’s inequality concerns correlations among measurements in a spatially entangled system, which is derived by assuming a local hidden-variable theory~\cite{Ref1.0}. The LGI is an analogous test based on temporal correlations of measurements of one observable of a system at different times~\cite{Ref1}. Both Bell's and Leggett-Garg inequalities are fulfilled within their \emph{classical} formulation; however, when the correlations are obtained using quantum states, these inequalities are violated and cannot be explained by classical theories. \\
LGI is based on the concept of \emph{macrorealism} (MR) and \emph{non-invasive measurability} (NIM). MR means that the system which has two or more macroscopically distinct states, pertaining to an observable $\hat{Q}$, always exists in one of these states, irrespective of any measurement performed on it. NIM states that we can measure $\hat{Q}$ without disturbing the future dynamics of the system. Given the two-time correlation function $C_{ij}=\langle Q(t_{i})Q(t_{j})\rangle$ for $t_j>t_i$, with a dichotomous observable $Q=\pm1$, the two simplest Leggett-Garg functions are given by~\cite{Ref1,Ref2}
\begin{equation}
\label{eq:1}
K_3=C_{12}+C_{23}-C_{13},
\end{equation}
\begin{equation}
\label{eq:2}
K_4=C_{12}+C_{23}+C_{34}-C_{14}.
\end{equation}
It has been established that $-3\leq K_3\leq 1$ and $-2\leq K_4\leq 2$, but, as discussed above, these inequalities are violated by quantum mechanics. \\

LGIs have been studied in many works, both theoretically and experimentally~\cite{ref1.1,ref1.2,ref1.3,ref1.4,ref1.5,ref1.6,ref1.7,ref1.8}. Several studies have focused on LGI violations in neutrino systems~\cite{ref1.9,ref1.10,ref1.11,ref1.12,ref1.13,ref1.14,Ref3,Ref4}. For the neutrino flavour oscillation in vacuum, as reported in~\cite{Ref3}, an LGI violation is found with a maximum value of $K_4=2.76$, with equal interval times and values of $\Delta m^2$ and the mixing angle $\theta$ from the KamLAND experimental setup. In addition, it has been found that the maximum value of $K_4$ depends sensitively on the mixing angle. In~\cite{Ref4}, it has been demonstrated how oscillation phenomena can be used to test violations of the classical limit by performing measurements on an ensemble of neutrinos at different energies, rather than on a single neutrino at different times. Here, the quantities $K_3$ and $K_4$ have been studied as a function of relative phases, yielding a prediction inconsistent with MR at the confidence level of $7\sigma$ for $K_4$, which constitutes a clear observation of LGI violations.\\
In this work, we study the LGI for neutrino flavour-and-spin oscillations in a constant magnetic field. This provides a suitable framework for exploring quantum aspects in relativistic particle systems in astrophysical contexts. We consider a neutron star, a system with an extreme astrophysical environment and a strong magnetic field.
\section{Massive Neutrino in a magnetic field}
\label{sec:1}
In the initial formulation of the Standard Model, neutrinos are massless particles with no electromagnetic properties. However, in an extension of this model, massive Dirac neutrinos have a non-zero magnetic moment, which implies that neutrinos have non-trivial electromagnetic properties~\cite{Ref5,Ref6}. The magnetic moment of the neutrino mass states is given by~\cite{Ref5,Ref6}
\begin{equation}\label{eq:3}
    \mu_{ii}^D=\frac{3eG_Fm_i}{8\sqrt{2}\pi^2}\approx 3.2\times10^{-19}\left(\frac{m_i}{1 \text{eV}}\right)\mu_B
\end{equation}
where $\mu_B$ is the Bohr magneton. As a consequence of the interaction of the neutrinos' magnetic moment with a transverse magnetic field, phenomena such as spin precession arise, and flavour oscillation patterns are modified. Spin precession causes left-handed neutrinos to become right-handed neutrinos and vice versa. A similar change occurs with flavour oscillation. This explains the four neutrino species that correspond to two different flavour states with both right-handed and left-handed helicities.
\subsection{Neutrino flavour and spin oscillations in a constant magnetic field.}
\label{sec:2}
The two neutrino flavour states are given by the equations:
\begin{eqnarray} 
\label{eq:2-1}
|\nu^{L(R)}_{e}\rangle&=&|\nu^{L(R)}_{1}\rangle\cos\theta+|\nu^{L(R)}_{2}\rangle\sin\theta,\\
\label{eq:2-2}
|\nu^{L(R)}_{\mu}\rangle&=&-|\nu^{L(R)}_{1}\rangle\sin\theta+|\nu^{L(R)}_{2}\rangle\cos\theta,
\end{eqnarray}
with $\nu^{L(R)}_{i}$ as the  neutrino mass states helicity ($i= 1, 2$). There are many ways to address neutrino oscillations in a magnetic field~\cite{Ref7,Ref8,Ref9,Ref10}. However, in this work, we adopt the approach implemented in~\cite{Ref11}. As helicity mass states $\nu^{L(R)}_{i}$ are not stationary in the presence of a magnetic field, we expand $\nu^{L(R)}_{i}$ over the neutrino stationary states $\nu^{-(+)}_{i}$. \\
Neutrino states $\nu^s_{i}$ ($s=\pm$1) of a massive neutrino that propagates along the $\textbf{n}_z$ direction in the presence of a constant
and arbitrarily oriented magnetic field can be found as the solution of the Dirac equation  
\begin{equation}
\label{eq:2-2a}
(\gamma^\mu p_\mu-m-\mu_{i}\boldsymbol{\Sigma}\cdot \textbf{B})|\nu^s_{i}(p)\rangle=0
\end{equation}
where $\mu_i$ is the neutrino magnetic moment and the magnetic
field is given by $\textbf{B}=(B_{\perp},0,B_{||})$. Equation~\eqref{eq:2-2a} can be re-written in the equivalent form
\begin{equation}
\label{eq:2-3a}
\hat{H}_i|\nu^s_{i}\rangle=E|\nu^s_{i}\rangle,
\end{equation}
where the Hamiltonian is
\begin{equation}
\label{eq:2-4a}
\hat{H}=\gamma_0\boldsymbol{\gamma}\cdot\textbf{p}+m\gamma_0+\mu_{i}\gamma_0\boldsymbol{\Sigma}\cdot\textbf{B}.
\end{equation}
The neutrino energy spectrum is given by 
\begin{equation}
\label{eq:2-5a}
E^s_i=\sqrt{m_i^2+p^2+\mu^2_{i}\textbf{B}^2+2\mu_is\sqrt{m_i^2\textbf{B}^2+p^2B^2_{\perp}}},
\end{equation}
where $s=\pm1$ corresponds to two different eigenvalues of the
Hamiltonian and $p=|\textbf{p}|$. For relativistic neutrino energies, we have $p \gg m$, and for realistic values of the neutrino magnetic moments and magnetic-fields strengths, we also obtain $p\gg \mu B$. With this approximation, the energy values are given by: 
\begin{equation}
\label{eq:2-5b}
E^s_i\approx p+\frac{m_i^2}{2p}+\frac{\mu^2_{i}B^2}{2p}+2\mu_i sB_{\perp}.
\end{equation}
The corresponding spin operator that commutes with the neutrino Hamiltonian in the magnetic field~(\ref{eq:2-4a})  can be chosen in the form~\cite{Ref11}:
\begin{equation}
\label{eq:2-5c}
\hat{S}_i=\frac{1}{N}\left[\boldsymbol{\Sigma}\cdot \textbf{B} -\frac{i}{m_i}\gamma_0\gamma_5(\boldsymbol{\Sigma}\times\textbf{p})\textbf{B}\right],
\end{equation}
with 
\begin{equation}
\label{eq:2-5f}
\frac{1}{N}=\frac{m_i}{\sqrt{m_i^2\textbf{B}^2+p^2B^2_{\perp}}}.
\end{equation}
For stationary neutrino states, the spin operator $\hat{S}_i$ has to satisfy the equation
\begin{eqnarray} 
\label{eq:2-5d}
\hat{S}_i|\nu^{s}_{i}\rangle&=&s|\nu^{s}_{i}\rangle,
\end{eqnarray}
with again, $s=\pm1$ and 
\begin{eqnarray} 
\label{eq:2-5e}
\langle\nu^{s}_{i}|\nu^{s'}_{j}\rangle&=&\delta_{ss'}\delta_{ij}.
\end{eqnarray}
A projector operator is introduced 
\begin{eqnarray} 
\label{eq:2-5ff}
\hat{P}^\pm_i=\frac{1\pm\hat{S}_i}{2}, 
\end{eqnarray}
where for stationary states,  $\hat{P}^\pm_i$ must satisfy the eigenvalue equation
\begin{eqnarray} 
\label{eq:2-5g}
\langle\nu^{s}_{i}|\hat{P}^\pm_i|\nu^{s'}_{j}\rangle&=&\delta_{ss'}\delta_{ij}.
\end{eqnarray}
Neutrino helicity states can be expanded over the neutrino stationary states
\begin{eqnarray} 
\label{eq:2-6}
|\nu_i^{L}(t)\rangle =
c_i^{+} e^{-iE_i^{+} t} |\nu_i^{+}\rangle
+
c_i^{-} e^{-iE_i^{-} t} |\nu_i^{-}\rangle,\\
\label{eq:2-7}
|\nu_i^{R}(t)\rangle =
d_i^{+} e^{-iE_i^{+} t} |\nu_i^{+}\rangle
+
d_i^{-} e^{-iE_i^{-} t} |\nu_i^{-}\rangle,
\end{eqnarray}
where $|\nu_i^\pm\rangle$ are stationary eigenstates of the Hamiltonian, and $c_i^\pm, d_i^\pm$ are time-independent coefficients determined from the initial conditions. Since ultra-relativistic neutrinos are produced in weak interaction process always as left-handed helicity states, in this approximation helicity and chiral states are almost indistinguishable.

The coefficients $c_i^\pm$ and $d^\pm_i$ are given by the matrix elements of the projector operators~(\ref{eq:2-5ff})
\begin{eqnarray} 
\label{eq:2-7b}
|c_i^\pm|^2&=&\langle\nu_i^L|\hat{P}^\pm_i|\nu_i^L\rangle =\frac{1}{2}\left(1\pm\frac{B_{||}}{N}\right), \\
\label{eq:2-7c}
|d_i^\pm|^2&=&\langle\nu_i^R|\hat{P}^\pm_i|\nu_i^R\rangle =\frac{1}{2}\left(1\mp\frac{B_{||}}{N}\right), \\
(d_i^\pm)^*c_i^\pm&=&\langle\nu_i^L|\hat{P}^\pm_i|\nu_i^R\rangle =\mp\frac{1}{2}\frac{p(B_{||}+iB_\perp)}{m_iN}, 
\end{eqnarray}
and, again $1/N$ is given by the Eq.~(\ref{eq:2-5f}). Neutrino stationary states are given by $|\nu^{s}_{i}(t)\rangle=e^{-iE^s_it}|\nu^{s}_{i}(0)\rangle$, from which the flavour states ~(\ref{eq:2-1}) and~(\ref{eq:2-2}) can be written as 
\begin{eqnarray} 
\label{eq:2-8}
\nonumber |\nu^{L}_{e}(t)\rangle&=&[c_1^+e^{-iE^+_1t}|\nu^{+}_{1}\rangle+c_1^-e^{-iE^-_1t}|\nu^{-}_{1}\rangle]\cos\theta \\
& &+[c_2^+e^{-iE^+_2t}|\nu^{+}_{2}\rangle+c_2^-e^{-iE^-_2t}|\nu^{-}_{2}\rangle]\sin\theta. \\
\label{eq:2-9}
\nonumber |\nu^{L}_{\mu}(t)\rangle&=&-[c_1^+e^{-iE^+_1t}|\nu^{+}_{1}\rangle+c_1^-e^{-iE^-_1t}|\nu^{-}_{1}\rangle]\sin\theta \\
& &+[c_2^+e^{-iE^+_2t}|\nu^{+}_{2}\rangle+c_2^-e^{-iE^-_2t}|\nu^{-}_{2}\rangle]\cos\theta,
\end{eqnarray}
where, for simplicity $|\nu^{\pm}_{1,2}\rangle=|\nu^{\pm}_{1,2}(0)\rangle$. The probability of flavour oscillation $\nu_{e}^L\leftrightarrow\nu_\mu^L$ is given by the expression
\begin{eqnarray}
\label{eq:2-10}
P_{\nu_\mu^L\rightarrow \nu_e^L}(t)&=&|\langle \nu^{L}_{\mu}(0)|\nu^{L}_{e}(t)\rangle|^2, 
\end{eqnarray}
which, with Eqs.~(\ref{eq:2-8}) and~(\ref{eq:2-9}) becomes 
\begin{eqnarray}
\label{eq:2-10a}
\nonumber P_{\nu_\mu^L\rightarrow \nu_e^L}(t)&=&\sin^2\theta\cos^2\theta\Big|-|c^+_1|^2e^{-iE^+_1t}-|c^-_1|^2e^{-iE^-_1t}\\
&& +|c^+_2|^2e^{-iE^+_2t}+|c^-_2|^2e^{-iE^-_2t}\Big|^2.
\end{eqnarray}
The oscillation probability in Eq.~(\ref{eq:2-10a}) has a set of characteristic energies. For simplicity, we consider the case where $\mu_1 = \mu_2 = \mu$. In the ultrarelativistic approximation ($p \gg m$), these characteristic energy values are:
\begin{eqnarray}
E^{+}_1 - E^{-}_1 &=& E^{+}_2 - E^{-}_2 = 2\mu B_\perp , \\
E^+_2 - E^+_1 &=&E^-_2 - E^-_1=\frac{\Delta m^2}{2p} ,\\
E^-_2 - E^+_1 &=& \frac{\Delta m^2}{2p}-2\mu B_\perp ,\\
E^+_2 - E^-_1 &=&\frac{\Delta m^2}{2p}+2\mu B_\perp .
\end{eqnarray}
The probability of flavour oscillation $P_{\nu_\mu^L\rightarrow \nu_e^L}(t)$ is simplified if one accounts for the relativistic neutrino energies $(p\gg m)$ and for realistic values of the neutrino magnetic moments and the strengths of magnetic fields $(p\gg\mu B)$. In this case, we
obtain from Eqs.~(\ref{eq:2-7b}) and~(\ref{eq:2-7c}) that $|c^\pm_i|=|d^\pm_i|\approx1/2$. Therefore, applying these conditions to the expression in Eq.~(\ref{eq:2-10a}), the probability of flavour oscillations $\nu_{\mu}^L\leftrightarrow\nu_e^L$ becomes: 
\begin{eqnarray}
\label{eq:2-11}
P_{\nu_\mu^L\rightarrow \nu_e^L}(t)&=&\cos^2(\mu B_\perp t)\sin^22\theta\sin^2\left(\frac{\Delta m^2}{4p}t\right). 
\end{eqnarray}
For completeness, the survival relation probability
\begin{eqnarray}
\nonumber P_{\nu_e^L\rightarrow \nu_e^L}(t)&=&|\langle \nu^{L}_{e}(0)|\nu^{L}_{e}(t)\rangle|^2\\
&=& \cos^2(\mu B_\perp t)\left[1-\sin^22\theta\sin^2\left(\frac{\Delta m^2}{4p}t\right)\right],
\label{eq:2-13}
\end{eqnarray}
is also calculated. The probability $P_{\nu_\mu^L\rightarrow \nu_\mu^L}$ is analogous to that in Eq.~(\ref{eq:2-13}). Notice that the probability of finding $\nu_\mu^L$ and $\nu_e^L$ at time $t$ is not $1$, i.e.,
\begin{equation}
\label{eq:2-14}
P_{\nu_\mu^L\rightarrow \nu_e^L}(t)+P_{\nu_\mu^L\rightarrow \nu_\mu^L}(t)=\cos^2(\mu B_\perp t).
\end{equation}
This result implies that there will be regions where the probability of finding $\nu_e^L$ and $\nu_\mu^L$ will be appreciable, i.e., when $P_{\nu_\mu^L\rightarrow \nu_e^L}(t)+P_{\nu_\mu^L\rightarrow \nu_\mu^L}(t)\approx 1$.\\
The spin oscillation $\nu_e^L\leftrightarrow\nu_e^R$ probability is given by
\begin{eqnarray}
\label{eq:2-15}
\nonumber P_{\nu_e^L\rightarrow \nu_e^R}(t)&=&|\langle \nu^{L}_{e}(0)|\nu^{R}_{e}(t)\rangle|^2\\
&=&\sin^2(\mu B_\perp t)\left[1-\sin^22\theta\sin^2\left(\frac{\Delta m^2}{4p}t\right)\right],
\end{eqnarray}
which are obtained in a similar way to the probability of flavour oscillation ~(\ref{eq:2-11}). 
\section{Leggett-Garg inequality for neutrino flavour and spin oscillations in a magnetic field}
\label{sec3}
The correlation function $C_{ij}$ is defined as~\cite{Ref2}
\begin{equation} 
\label{eq:3-1}
C_{ij}=\sum_{Q_i,Q_j=\pm 1}Q_iQ_jP_{ij}(Q_i,Q_j).
\end{equation}
Here $P_{ij}(Q_i,Q_j)$ represents the joint probability of measuring $Q_i = \pm1$ at time $t_i$ and $Q_j = \pm1$ at time $t_j$. Most systems in which LGI has been tested use a single observable, but here we can test inequalities using two dichotomous observables, i.e., flavour  and spin. We study e ach case separately here.
\subsection{LGI for neutrino flavour oscillation.}
\label{sec3-1}
Let us consider neutrino flavour oscillations in a magnetic field, taking the main system $\nu_e^L\to\nu_\mu^L$. We take $Q=+1$  when the system is in the $|\nu_{e}^L\rangle$ state and $Q=-1$ when the system is in the $|\nu_{\mu}^L\rangle$ state.
The correlation function $C_{12}$ given by Eq.~(\ref{eq:3-1}) can be expressed in terms of joint probabilities using the notation introduced in~\cite{Ref3}. If $|\nu_\mu^L\rangle$ is the initial state, we write the joint probability as
\begin{equation}
P_{\nu_\mu^L\to\nu_e^L}(t_1)P_{\nu_e^L\to\nu_e^L}(t_2)=P_{\nu_e^L,\nu_e^L}(t_1,t_2),     
\end{equation}
and thus we have the correlation function in terms of four joint probabilities, given by the expression 
\begin{eqnarray}
 \label{eq:3-4}
 C_{12}&=&P_{\nu_e^L,\nu_e^L}(t_1,t_2)-P_{\nu_e^L,\nu_\mu^L}(t_1,t_2)-P_{\nu_\mu^L,\nu_e^L}(t_1,t_2)+ P_{\nu_\mu^L,\nu_\mu^L}(t_1,t_2).
\end{eqnarray}
Again, in this context $P_{\nu_e^L,\nu_e^L}(t_1,t_2)$ denotes the joint probability of finding the state $|\nu_{e}^L\rangle$ at $t_1$ and $t_2$, and $P_{\nu_e^L,\nu_\mu^L}(t_1,t_2)$ is the joint probability of finding the states $|\nu_{e}^L\rangle$, $|\nu_{\mu}^L\rangle$ at the respective times $t_1$ and $t_2$, etc.\\
The flavour oscillation probabilities are given by Eqs.~(\ref{eq:2-11}) and~(\ref{eq:2-13}). If $|\nu_\mu^L\rangle$ is the initial state, the joint probability of finding the state $|\nu_{e}^L\rangle$ at $t_1$ and $t_2$ is
\begin{eqnarray}
\label{eq:3-5}
\nonumber P_{\nu_e^L,\nu_e^L}(t_1,t_2)&=& \sin^22\theta\sin^2\left(\frac{\Delta m^2}{4p}t_1\right)\cos^2[\mu B_\perp t_1] \\
&& \times \left[1-\sin^2 2\theta\sin^2\left(\frac{\Delta m^2}{4p}\Delta t\right)\right]\cos^2(\mu B_\perp\Delta t),
\end{eqnarray}
where $\Delta t=t_2-t_1$.
The remaining three joint probabilities can be calculated in a similar way. The above allows us to determine the correlation function $C_{12}$:
\begin{eqnarray}
\label{eq:3-6}
C_{12}&=&\cos^2(\mu B_\perp t_1)\cos^2(\mu B_\perp \Delta t)\left[1-2\sin^22\theta\sin^2\left(\frac{\Delta m^2}{4p}\Delta t\right)\right].
\end{eqnarray}
Notice that $C_{12}$ no longer depends on $t_1$ when $B_\perp=0$, but there can still be values of $\Delta t$ for which $C_{12}\neq 0$. The correlation functions $C_{23}$, $C_{34}$, $C_{13}$, and $C_{14}$ are calculated similarly to Eq.~(\ref{eq:3-6}), as in the calculation of the functions $K_3$ and $K_4$ in Eqs.~(\ref{eq:1}) and~(\ref{eq:2}). We assume equal time intervals between measurements for all correlation functions, i.e., $t_4-t_3=t_3-t_2=t_2-t_1=\Delta t$. This gives $t_2=t_1+\Delta t$ and $t_3=t_1+2\Delta t$, which is important for calculating $C_{ij}$ and subsequently the $K_3$ and $K_4$ functions. They are written in terms of a distance function using the relativistic limit $t \simeq L$. Therefore, $L_1$ and $\Delta L$ represent the distances traveled by the neutrinos in time $t_1$ and interval $\Delta t$, respectively.The final results for the functions $K_3$ and $K_4$ of the neutrino flavour oscillation are
\begin{eqnarray}
\label{eq:3-7a}
\nonumber K_3&=&\cos^2(\mu B_\perp\Delta L)\bigg[\cos^2(\mu B_\perp L_1)+\cos^2\{\mu B_\perp(L_1+\Delta L)\}\bigg]\\
\nonumber&&\times\left[1-2\sin^22\theta \sin^2\left( \frac{\Delta m^2}{4p}\Delta L\right)\right]\\
\nonumber & &-\cos^2(\mu BL_1)\cos^2(2\mu B_\perp\Delta L)\left[1-2\sin^22\theta \sin^2\left(\frac{\Delta m^2}{4p}2\Delta L\right)\right]\\.
\end{eqnarray}
\begin{eqnarray}
\label{eq:3-7}
\nonumber K_4&=&\bigg[\cos^2(\mu B_\perp L_1)+\cos^2\{\mu B_\perp(L_1+\Delta L)\}
+\cos^2\{\mu B_\perp(L_1+2\Delta L)\}\bigg]\\
\nonumber& &\times\cos^2(\mu B_\perp\Delta L)\left[1-2\sin^22\theta \sin^2 \left(\frac{\Delta m^2}{4p}\Delta L\right)\right] \\
\nonumber & &-\cos^2(\mu B_\perp L_1)\cos^2(3\mu B_\perp\Delta L)
\left[1-2\sin^22\theta \sin^2 \left(\frac{\Delta m^2}{4p}3\Delta L\right)\right].\\
\end{eqnarray}
As special cases, when $B_\perp=0$ the Eq.~(\ref{eq:3-7}) reduces to 
\begin{eqnarray}
    \label{eq:3-8}
K_4=2-2\sin^22\theta\bigg[3\sin^2\left(\frac{\Delta m^2}{4p}\Delta L\right)
-\sin^2\left(\frac{\Delta m^2}{4p}3\Delta L\right)\bigg],
\end{eqnarray}
which is the usual LGI for vacuum neutrino flavour oscillation studied in~\cite{Ref3}.
The experimental values of $\Delta m^2$ and $\theta$, taken from the KamLAND experiment, are $7.58\times10^{-5}$ eV$^2$ and $36.80^\circ$, respectively~\cite{Ref12}. The Standard Model prediction~(\ref{eq:3}) for the magnetic moment is $\mu \sim 10^{-19}\mu_B$ for a neutrino mass of $m = 0.5$ eV.
\begin{figure}[ht]
\centerline{%
\includegraphics[width=12.5cm]{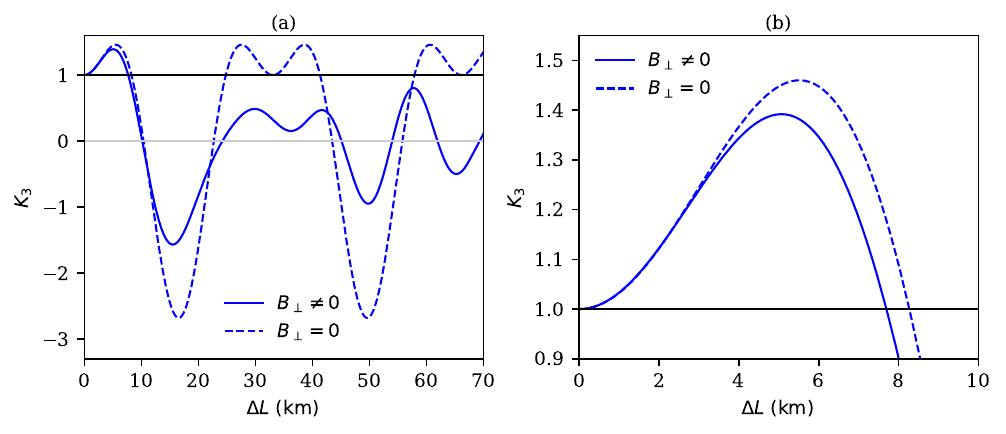}}
\caption{(a) \small{$K_3$ plot for the flavour oscillation in a transverse magnetic field $B=10^{12}$ T, a neutrino energy $p=1$ MeV, $\Delta m^2=7.58\times10^{-5}$ eV$^2$, a magnetic moment $\mu=10^{-19}\mu_B$ and a mixing angle $\theta=36.80^\circ$. Horizontal black line denotes the maximum value of $K_3$ allowed by LGI. (b) $K_3$ for the flavour oscillation within a smaller length region where $K_3>1$, $B_\perp=0$, and $B_\perp\neq0$.}}
\label{fig:1.1}
\end{figure}
\begin{figure}[ht]
\centerline{%
\includegraphics[width=12.5cm]{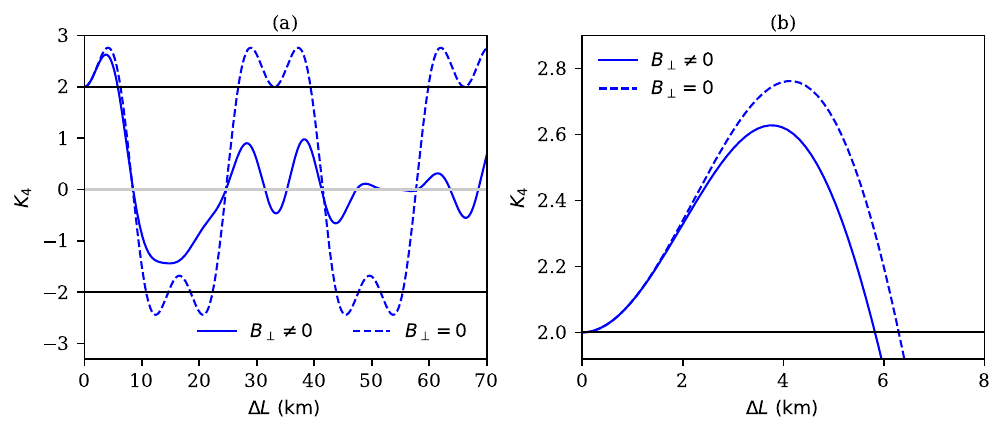}}
\caption{\small{(a) Plot of $K_4$ for the flavour oscillation in a transverse magnetic field $B=10^{12}$ T, a neutrino energy $p=1$ MeV, $\Delta m^2=7.58\times10^{-5}$ eV$^2$, a magnetic moment $\mu=10^{-19}\mu_B$ and a mixing angle $\theta=36.80^\circ$. Horizontal black lines denote the maximum and minimum values of $K_4$ allowed by LGI. (b) $K_4$ for the flavour oscillation within a smaller length region where $K_4>2$, $B_\perp=0$, and $B_\perp\neq0$.}}
\label{fig:1}
\end{figure}
Several options for $L_1$ can be chosen; however, we find that LGI violation occurs only when $L_1$ lies within length intervals where the probability of oscillation is appreciable. This is given by the oscillation length $L+\Delta L=1/(\mu B)$. For a neutron star, such as magnetar~\cite{Ref13}, whose size is $L\sim 20$ km and magnetic field values $B\sim10^{12}-10^{15}$ T, we can extend our study of LGIs to this kind of system.  In Figs.~\ref{fig:1.1}(a) and~\ref{fig:1}(a), we show the plots of the $K_3$ and $K_4$ functions for neutrino flavour oscillations in a transverse magnetic field with $B_\perp = 10^{12}$ T, a distance $L_1 = 0$ km, and neutrino energy $p = 1$ MeV. These values generate a length oscillation $\Delta L=1/(\mu B)\sim 30$ km, which is of the order of a magnetar's size. Both functions display regions where flavour oscillations lead to violations of the LGI. In Figs.~\ref{fig:1.1}(b) and~\ref{fig:1}(b), we show plots for $K_3$ and $K_4$ functions for cases with and without magnetic field, but within a smaller length region, i.e., a region of size comparable to the neutron star where LGIs are violated. We observe an interesting behavior of the LGIs when the magnetic field is nonzero: there are values of $\Delta L$ where the LGI is not violated when $B \neq 0$; however, in these same regions, the LGI is violated when $B = 0$.\\
Within the plotted range of $\Delta L$, a maximum peak appears in both $K_3$ and $K_4$. Specifically, the maximum values attained within the length region from $0$ to $70$ km are $K_3 = 1.39$ and $K_4 = 2.62$, occurring at $\Delta L = 5.06$ km and $\Delta L = 3.76$ km, respectively. Notably, all these regions in which LGIs are violated lie within the magnetar's size order of magnitude. For comparison, in the zero-field case, the maximum values reached are $K_3 = 1.46$ and $K_4 = 2.76$. In addition, LGIs are violated over larger length intervals. For instance, we obtain $K_3 = 1.42$ at $\Delta L = 104.80$ km and $K_4 = 2.52$ at $\Delta L = 103.50$ km; however, these lengths are significantly larger than the magnetar size.
\\
Let us consider the case where $\Delta m^2 = 0$. In this scenario, there will obviously be no flavour oscillation, and Eq.~(\ref{eq:3-8}), when the field is zero, indicates that we remain in the classical regime, with $K_4 = 2$. However, if we instead set $\Delta m^2 = 0$ in Eqs.~(\ref{eq:3-7a}) and~(\ref{eq:3-7}), where the magnetic field remains nonzero, an LGI violation still persists with maximum values of $K_3=1.08$ for $\Delta L=14.50$ km and $K_4=2.02$ for $\Delta L=6.46$ km. In~\cite{ref1.12}, where the LGI for three-flavour neutrino oscillations was studied, the authors also examined the case $\Delta m^2 = 0$ and found an LGI violation. This result can be explained by the fact that, for three-state neutrino oscillations, there are three neutrino mass states, and if two of them become equal, then there will be the possibility of neutrino oscillations because there are now effectively two masses. In our case, the violation of the LGI when $\Delta m^2=0$ is explained by the fact that although the flavour transition probability is zero, the survival probability is still non-zero. This indicates that, if we consider $|\nu_\mu^L\rangle$ as the initial state, the probability of detecting $|\nu_e^L\rangle$ is zero. However, the probability of finding $|\nu_\mu^L\rangle$ is not equal to 1, as stated in Eqs.~(\ref{eq:2-11}) and~(\ref{eq:2-13}). This implies that measurements of the initial state $|\nu_\mu^L\rangle$ are not trivial. Thus, LGI violation when $\Delta m^2=0$ can be clearly attributed to the influence of the magnetic field on the probabilities of flavour oscillation.\\
Although the present analysis has been restricted to the two-flavour approximation, it is important to note how the situation changes when the full three-flavour framework is considered. The behavior of three-flavour oscillations in a magnetic field reproduces the qualitative effects observed in the two-flavour case~\cite{Ref14}; namely, the magnetic field modulates the vacuum oscillation probabilities. However, both two- and three-flavour systems possess their own dynamical structure, so their quantitative evolution differs, even though the qualitative behavior of $K_3$ and $K_4$ remains similar. Moreover, in the three-flavour case, the mixing matrix involves three mixing angles and one complex phase that introduces possible CP-violating effects, which may modify the temporal correlations relevant to the LGI. For instance, Ref.~\cite{ref1.12} reported that the presence of the CP-violating phase $\delta_{\mathrm{CP}}$ can enhance the maximum violation of the LGI, indicating that CP violation may strengthen the quantum nature of the three-flavour neutrino oscillation system~\cite{ref1.12}.
\subsection{LGI for neutrino spin oscillation.}
\label{sec3-2}
\begin{figure}[ht]
\centerline{%
\includegraphics[width=12.5cm]{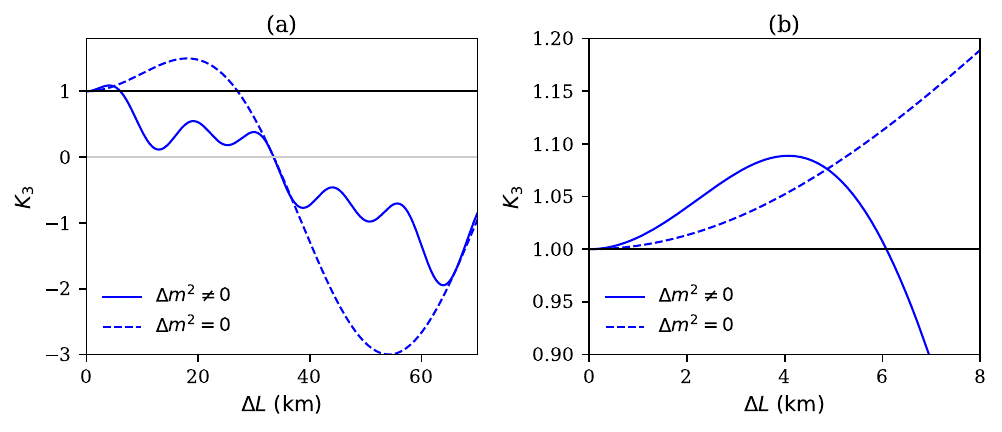}}
\caption{\small{(a) $K_3$ plot for the spin oscillation in a transverse magnetic field $B=10^{12}$ T, a neutrino energy $p=1$ MeV, $\Delta m^2=7.58\times10^{-5}$ eV$^2$, a magnetic moment $\mu=10^{-19}\mu_B$ and a mixing angle $\theta=36.80^\circ$. Horizontal black line denotes the maximum value of $K_3$ allowed by LGI. (b) $K_3$ for the spin oscillation with a smaller length region where $K_3>1$, $\Delta m=0$ and $\Delta m\neq0$.}}
\label{fig:2.1}
\end{figure}

\begin{figure}[ht]
\centerline{%
\includegraphics[width=12.5cm]{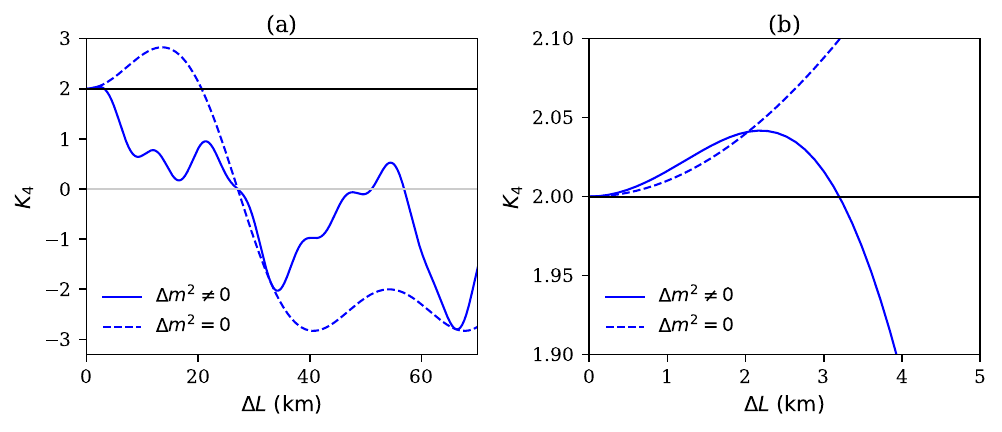}}
\caption{\small{(a) $K_4$ plot for the spin oscillation in a transverse magnetic field $B=10^{12}$ T, a neutrino energy $p=1$ MeV, $\Delta m^2=7.58\times10^{-5}$ eV$^2$, a magnetic moment $\mu=10^{-19}\mu_B$ and a mixing angle $\theta=36.80^\circ$. Horizontal black lines denote the maximum and minimum values of $K_4$. (b) $K_4$ for spin oscillation within a smaller length region where $K_4>2$, $\Delta m=0$ and $\Delta m\neq0$.}}
\label{fig:2}
\end{figure}

For neutrino spin oscillations in a magnetic field, we take the main system to be the $\nu_e^L\to\nu_e^R$ system. The spin oscillation probabilities are given by Eqs.~(\ref{eq:2-15}) and~(\ref{eq:2-13}).  
We assign $Q = +1$ when the system is in the $|\nu_e^R\rangle$ state and $Q = -1$ when it is in the $|\nu_e^L\rangle$ state. In the same way as described in Sect.~\ref{sec3-1}, the correlation function $C_{12}$ is expressed in terms of the four joint probabilities $P_{\nu_e^R, \nu_e^R}(t_1, t_2)$, $P_{\nu_e^R, \nu_e^L}(t_1, t_2)$, $P_{\nu_e^L, \nu_e^R}(t_1, t_2)$, and $P_{\nu_e^L, \nu_e^L}(t_1, t_2)$, which can be calculated similarly to the case of flavour oscillations. Therefore, the time-dependent function $C_{12}$ obtained in this case is given by
\begin{eqnarray}
\label{eq:3-11}
\nonumber C_{12}&=&\left(1-2\sin^2(\mu B_\perp\Delta L)\right)\left[1-\sin^22\theta\sin^2\left(\frac{\Delta m^2}{4p}L_1\right)\right]\\
& &\times\left[1-\sin^22\theta\sin^2\left(\frac{\Delta m^2}{4p}\Delta L\right)\right].
\end{eqnarray}
Applying the same procedure described in Sect.~\ref{sec3-1}, the functions $K_3$ and $K_4$ are obtained straightforwardly, and their corresponding results are
\begin{eqnarray}
\label{eq:3-13}
\nonumber K_3 &=& \left[2 - \sin^2 2\theta \sin^2\left(\frac{\Delta m^2}{4p}L_1\right) 
- \sin^2 2\theta \sin^2\left(\frac{\Delta m^2}{4p} (L_1 + \Delta L) \right) \right]\\ 
&&\times\left(1 - 2\sin^2(\mu B_\perp \Delta L)\right)\left[1 - \sin^2 2\theta \sin^2 \left( \frac{\Delta m^2}{4p} \Delta L \right)\right] \nonumber \\
\nonumber  & &- \left(1 - 2\sin^2(2\mu B_\perp \Delta L)\right) 
\left[1 - \sin^2 2\theta \sin^2 \left( \frac{\Delta m^2}{4p} L_1 \right)\right]\\
& &\times\bigg[1 - \sin^2 2\theta \sin^2 \left( \frac{\Delta m^2}{4p} 2\Delta L \right)\bigg].
\end{eqnarray}

\begin{eqnarray}
\label{eq:3-14}
\nonumber K_4 &=& \left(1 - 2\sin^2(\mu B_\perp \Delta L)\right){\left[1 - \sin^2 2\theta \sin^2 \left(\frac{\Delta m^2}{4p} \Delta L\right)\right]}\\
\nonumber && \bigg[3 - \sin^2 2\theta \sin^2\left(\frac{\Delta m^2}{4p}L_1\right) 
- \sin^2 2\theta \sin^2\left(\frac{\Delta m^2}{4p} (L_1 + \Delta L)\right) \\
\nonumber & &- \sin^2 2\theta \sin^2\left(\frac{\Delta m^2}{4p} (L_1 + 2\Delta L)\right)\bigg] \\
\nonumber &&-\left(1 - 2\sin^2(\mu B_\perp 3\Delta L)\right)\left[1 - \sin^2 2\theta \sin^2 \left(\frac{\Delta m^2}{4p} 3\Delta L\right)\right] \\
&&\times {\left[1 - \sin^2 2\theta \sin^2 \left(\frac{\Delta m^2}{4p} L_1\right)\right]}.
\end{eqnarray}
In Figs.~\ref{fig:2.1}-\ref{fig:2}, we show the plots of the $K_3$ and $K_4$ functions for neutrino spin oscillations using the same values as in Sec.(\ref{sec3-1}) for $p$, $B$, $\mu$ and $\Delta m^2$. Regions where the LGI is violated can be identified. The maximum LGI violation values obtained in the plotted region are $K_3 = 1.08$ at $\Delta L = 4.02$~km and $K_4 = 2.08$ at $\Delta L = 2.03$~km. Moreover, the LGI is violated over larger length intervals; for example, we obtain $K_3 = 1.28$ for $\Delta L = 97.51$~km and $K_4 = 2.63$ for $\Delta L = 98.62$~km. 
As a special case, we consider the situation where $\Delta m^2 = 0$. Here again, the LGI violation persists alongside the flavour oscillations, as can be seen in Figs.~\ref{fig:2.1}(a) and~\ref{fig:2}(a). A maximum LGI violation occurs at $K_3 = 1.5$ and $K_4 = 2.82$.
Finally, we consider the case where we set $B_{\perp}=0$ in  Eqs.~(\ref{eq:3-13}) and~(\ref{eq:3-14}) for $K_3$ and $K_4$. Here we get a violation of the LGI, with, \emph{e.g.}, $K_3=1.06$ for $\Delta L=4.29$ km and $K_4=2.01$ for $\Delta L=1.73$ km. Again, this situation is analogous to flavour oscillations with $\Delta m^2 = 0$, with the distinction that in this case the violation arises from  the dependence of the spin–oscillation probabilities on $\Delta m^2$.
\section{Conclusions}
\label{sec:4}
We studied the LGIs for the cases of neutrino flavour oscillations, $\nu_e^L \leftrightarrow \nu_\mu^L$, and spin oscillations, $\nu_e^L \leftrightarrow \nu_e^R$, in the presence of a constant magnetic field through the $K_3$ and $K_4$ functions. We determined that these functions exhibit $\Delta L$ regions where LGI violations occur. For flavour oscillations, we found maximum values of $K_3 = 1.39$ and $K_4 = 2.62$, whereas for spin oscillations we obtained $K_3 = 1.08$ and $K_4 = 2.08$. As special cases, we also studied the regime with $\Delta m^2 = 0$ and again identified regions where LGI violations persist, finding maximum values of $K_3 = 1.08$ and $K_4 = 2.02$ for flavour oscillations, and $K_3 = 1.5$ and $K_4 = 2.82$ for spin oscillations. It is worth highlighting that all the scenarios investigated here correspond to systems with strong magnetic fields, such as the extreme astrophysical environments discussed in~\cite{Ref11}. In this context, magnetars are systems to which the results obtained here for neutrino oscillations can be applied. Finally, in the case of a vanishing magnetic field, we also identified $\Delta L$ intervals where the $K_3$ and $K_4$ functions violate the LGIs for flavour oscillations (as studied in~\cite{Ref4}), and additionally for spin oscillations—an aspect less commonly explored in neutrino theory.

\bibliographystyle{unsrt}

\begin{thebibliography}{}

\bibitem{Ref1.0}
J. S. Bell, \href{https://journals.aps.org/ppf/abstract/10.1103/PhysicsPhysiqueFizika.1.195}{\textit{Physics} \textbf{1}, 195 (1964)}
\bibitem{Ref1}
A. J. Leggett, A. Garg, \href{https://journals.aps.org/prl/abstract/10.1103/PhysRevLett.54.857}{\textit{Phys. Rev. Lett.} \textbf{54}, 857(1985)}
\bibitem{Ref2}
Clive Emary \textit{et al},   \href{https://iopscience.iop.org/article/10.1088/0034-4885/77/1/016001}{\textit{Rep.Prog.Phys.} \textbf{77} 039501 (2014)}
\bibitem{ref1.1}
J. Naikoo, A. K. Alok, S. Banerjee, \href{https://journals.aps.org/prd/abstract/10.1103/PhysRevD.97.053008}{\textit{Phys. Rev. D} \textbf{97}, 053008 (2018)}.
\bibitem{ref1.2}
K. Sharma, A. Mahapatra, P. K. Panigrahi, S. Patra, \href{https://link.springer.com/article/10.1140/epjp/s13360-024-05666-2}{\textit{
Eur. Phys. J. Plus} \textbf{139}, 1-17 (2024)}.
\bibitem{ref1.3}
L. Rosales-Zárate, B. Opanchuk, Q. Y. He, M. D. Reid, \href{https://journals.aps.org/pra/abstract/10.1103/PhysRevA.97.042114}{\textit{Phys. Rev. A} \textbf{97}, 042114 (2018)}.
\bibitem{ref1.4}
N. Lambert, R. Johansson, F. Nori, \href{https://journals.aps.org/prb/abstract/10.1103/PhysRevB.84.245421}{\textit{Phys. Rev. B} \textbf{84}, 245421 (2011)}.
\bibitem{ref1.5}
C. Budroni, C. Emary, \href{https://journals.aps.org/prl/abstract/10.1103/PhysRevLett.113.050401}{\textit{Phys. Rev. Lett.} \textbf{113}, 050401 (2014)}.
\bibitem{ref1.6}
M. E. Goggin, M. P. Almeida, M. Barbieri, B. P. Lanyon, J. L. O’Brien, A. G. White, G. J. \href{https://www.pnas.org/doi/10.1073/pnas.1005774108}{\textit{Pryde, Proc. Natl. Acad. Sci. U.S.A.} \textbf{108}, 1256-1261 (2011)}.
\bibitem{ref1.7}
G. C. Knee, S. Simmons, E. M. Gauger, J. J. Morton, H. Riemann, N. V. Abrosimov, S. C. Benjamin, \href{https://www-nature-com.translate.goog/articles/ncomms1614?error=cookies_not_supported&code=e9308e47-dc0f-4f06-a4c9-be119b56ff2d&_x_tr_sl=en&_x_tr_tl=es&_x_tr_hl=es&_x_tr_pto=tc}{\textit{Nat. Commun.} \textbf{3}, 606 (2012)}.
\bibitem{ref1.8}
H. Katiyar, A. Brodutch, D. Lu, R. Laflamme, \href{https://iopscience.iop.org/article/10.1088/1367-2630/aa5c51/meta}{\textit{New J. Phys.} \textbf{19}, 023033 (2017)}.
\bibitem{ref1.9}
J. S. Xu, C. F. Li, X. B. Zou, G. C. Guo, \href{https://www.nature.com/articles/srep00101}{\textit{Sci. Rep.} \textbf{1}, 101 (2011)}.
\bibitem{ref1.10}
M. Blasone, F. Illuminati, L. Petruzziello, L. Smaldone, \href{https://journals.aps.org/pra/abstract/10.1103/PhysRevA.108.032210}{\textit{Phys. Rev. A} \textbf{108}, 032210 (2023)}.
\bibitem{ref1.11}
X. Z. Wang, B. Q. Ma, \href{https://link.springer.com/article/10.1140/epjc/s10052-022-10053-1}{\textit{Eur. Phys. J. C} \textbf{82}, 133 (2022)}.
\bibitem{ref1.12}
D. Gangopadhyay, A. S. Roy, \href{https://link.springer.com/article/10.1140/epjc/s10052-017-4837-2#citeas}{\textit{Eur. Phys. J. C} \textbf{77}, 260 (2017).}.

\bibitem{ref1.13}
Q. Fu, X. Chen, 
\href{https://link.springer.com/article/10.1140/epjc/s10052-017-5371-y}
{\textit{Eur. Phys. J. C} \textbf{77}, 1-6 (2017).}
\bibitem{ref1.14}
S. Shafaq, T. Kushwaha, P. Mehta, \href{https://arxiv.org/abs/2112.12726}{\emph{arXiv}:2112.12726 (2021)}.
\bibitem{Ref3}
D. Gangopadhyay, D. Home, A. SinhaRoy, \href{https://journals.aps.org/pra/abstract/10.1103/PhysRevA.88.022115}{\emph{Phys.Rev.A} \textbf{88}, 022115 (2013)}
\bibitem{Ref4}
J. A. Formaggio, D. I. Kaiser, M. M. Murskyj, T. E. Weiss, \href{https://journals.aps.org/prl/abstract/10.1103/PhysRevLett.117.050402}{\emph{Phys.Rev.Lett.} \textbf{117}, 050402 (2016)}
\bibitem{Ref5}
K. Fujikawa, R. Shrock, \href{https://journals.aps.org/prl/abstract/10.1103/PhysRevLett.45.963}{\emph{Phys. Rev. Lett.} \textbf{45}, 963 (1980)}
\bibitem{Ref6}
A. Studenikin, \href{https://pos.sissa.it/314/137/pdf}{\emph{PoS EPS–HEP2017}, \textbf{137} (2017)}.
\bibitem{Ref7}
J. Schechter, J.W.F. Valle, \href{https://journals.aps.org/prd/abstract/10.1103/PhysRevD.24.1883}{\emph{Phys. Rev. D} \textbf{24}, 1883 (1981)}
\bibitem{Ref8}
L. Okun, M. Voloshin, M. Vysotsky, \href{http://jetp.ras.ru/cgi-bin/dn/e_064_03_0446.pdf}{\emph{Sov. Phys. JETP} \textbf{64}, 446 (1986)}
\bibitem{Ref9}
E. Akhmedov, \href{https://www.sciencedirect.com/science/article/abs/pii/0370269388910489}{\emph{Phys. Lett. B} \textbf{213}, 64 (1988)}
\bibitem{Ref10}
C.-S. Lim, W. Marciano, \href{https://journals.aps.org/prd/abstract/10.1103/PhysRevD.37.1368}{\emph{Phys. Rev. D} \textbf{37}, 1368 (1988)}
\bibitem{Ref11}
A. Popov, A. Studenikin, \href{https://link.springer.com/article/10.1140/epjc/s10052-019-6657-z}{\emph{Eur. Phys. J. C} \textbf{79}, 144 (2019)}.
\bibitem{Ref12}
S. Abe, et al., \href{https://journals.aps.org/prl/abstract/10.1103/PhysRevLett.100.221803}{\emph{Phys. Rev. Lett.} \textbf{100}, 221803 (2008)}.
\bibitem{Ref13}
Victoria M. Kaspi, Andrei Beloborodov, \href{https://www.annualreviews.org/content/journals/10.1146/annurev-astro-081915-023329}{\emph{Annual Review of Astronomy and Astrophysics} \textbf{55}: 261-301 (2017)}.
\bibitem{Ref14}
Lichkunov, A., Popov, A., Studenikin, A.,
\href{https://arxiv.org/abs/2207.12285}{\emph{arXiv}:	arXiv:2207.12285 (2022)}.
\end{thebibliography}

\end{document}